\documentclass[sigconf]{acmart}
\AtBeginDocument{%
  \providecommand\BibTeX{{%
    \normalfont B\kern-0.5em{\scshape i\kern-0.25em b}\kern-0.8em\TeX}}}

\copyrightyear{2020}
\acmYear{2020}
\setcopyright{acmlicensed}
\acmConference[PARMA-DITAM'20]{11th Workshop on Parallel Programming
and Run-Time Management Techniques for Many-core Architectures / 9th
Workshop on Design Tools and Architectures for Multicore Embedded
Computing Platforms}{January 21, 2020}{Bologna, Italy}
\acmBooktitle{11th Workshop on Parallel Programming and Run-Time
Management Techniques for Many-core Architectures / 9th Workshop on
Design Tools and Architectures for Multicore Embedded Computing Platforms
(PARMA-DITAM'20), January 21, 2020, Bologna, Italy}
\acmPrice{15.00}
\acmDOI{10.1145/3381427.3381429}
\acmISBN{978-1-4503-7545-0/20/01}

\usepackage{breqn}
\usepackage{multirow}

\usepackage[skip=0pt]{caption}

\usepackage{float}

\usepackage{textgreek}

\usepackage{amsmath,array,graphicx}
\usepackage{kantlipsum}
\usepackage[export]{adjustbox}
\usepackage{enumitem}

\usepackage{array}
\usepackage{makecell}

\title{Run-Time Power Modelling in Embedded GPUs with Dynamic Voltage and Frequency Scaling}

\author{Jose Nunez-Yanez, Kris Nikov, Kerstin Eder, Mohammad Hosseinabady}
\email{{j.l.nunez-yanez,kris.nikov,kerstin.eder}@bristol.ac.uk,  mohammad@hosseinabady.com}
\affiliation{%
  \institution{Department of Electrical and Electronic Engineering,}
  \city{Bristol}
}

\begin{document}

\begin{abstract}
This paper investigates the application of a robust CPU-based power modelling methodology that performs an automatic search of explanatory events derived from performance counters to embedded GPUs. A 64-bit Tegra TX1 SoC is configured with DVFS enabled and multiple CUDA benchmarks are used to train and test models optimized for each frequency and voltage point. These optimized models are then compared with a simpler unified model that uses a single set of model coefficients for all frequency and voltage points of interest. To obtain this unified model, a number of experiments are conducted to extract information on idle, clock and static power to derive power usage from a single reference equation. The results show that the unified model offers competitive accuracy  with an average 5\% error with four explanatory variables on the test data set and it is capable to correctly predict the impact of voltage, frequency and temperature on power consumption. This model could be used to replace direct power measurements when these are not available due to hardware limitations or worst-case analysis in emulation platforms. 

\end{abstract}


\maketitle
\section{Introduction}

Embedded GPUs (Graphical Processing Units) which are physically present in the same chip as the central processing unit (CPU) are popular as general-purpose accelerators in power constrained applications such as unmanned aerial vehicles (UAV), self-driving cars or robotics. 

The power profiles of these GPUs are in order of Watts compared with the hundreds of Watts needed in their desktop counterparts that invariably use the PCIe bus to communicate with the host CPU and memory.  Power optimization in these embedded devices is of critical importance since in these applications, the power sources tend to be batteries and the systems must operate untethered for as long as possible. In this paper we investigate the application of the power modelling framework created in~\cite{Nikov2020} for heterogeneous embedded CPUs to embedded GPUs capable of general-purpose computing thanks to their support for languages such as as CUDA and OpenCL. Our framework, called ROSE (RObust Statistical search of explanatory activity Events), can be used to automatically collect activity and power data and then perform a complete search for optimal events across a large range of frequency and voltage pairs as defined in the device DVFS (Dynamic Voltage and Frequency Scaling) tables. ROSE multiple linear regression optimization uses OLS (Ordinary Least Squares) which is well understood for the case of both desktop and integrated CPUs but it has been less studied in GPUs that are characterized by a proprietary black box architecture with restricted access to internal microarchitecture details. Taking these points into account, the novelty of this work can be summarized as follows:  
\begin{enumerate}[align=left]
\item[1] We perform power modelling on an embedded GPU device with integrated power measurement and voltage/frequency scaling compared with previous work largely focused on desktop GPUs.
\item[2] We limit the number of explanatory variables used in the model to enable the run-time collection of this information using a limited number of hardware registers. 
\item[3] We propose a novel unified model that includes temperature, frequency and voltage as global states and compare it against models with coefficients optimized for single temperatures and frequency/voltage pairs.
\end{enumerate}

This paper is organized as follows. Section 2 introduces related work in the area of power modelling with GPUs. 
Section 3 presents the methodology based on our previous work in this area, the set of CUDA benchmarks for model training/verification and the techniques used to obtain run-time measures of power and event information. Section 4 develops models based on this methodology with coefficients optimized for individual voltage and frequency pairs.  Section 5 proposes an unified model so that the power of a single per-frequency model can be scaled to an extended range of voltage and frequency points. Section 6 investigates temperature effects on power consumption and model accuracy. Finally, section 7 concludes the paper.

\section{Background}

As previously indicated, there is a significant amount of work on power modelling of CPU cores and CPU-based systems and the interested reader is referred to ~\cite{nunez} for a review of results and techniques. In the field of general-purpose GPU-based computing, the amount of power modelling research is more limited. The authors of ~\cite{nag10} investigate how performance counters can be used to model power on a desktop NVIDIA GPUs connected to a host computer via a PCIe interface. The PCIe interface is instrumented with current clamp sensors and the host computer samples these sensors while collecting performance counter information. The authors identify a total of 13 CUDA performance counters but since only four hardware registers are available, multiple runs are needed to access all of them. The authors also identify that certain kernels that perform texture reads, such as Rodinia Leukocyte, show significant power error up to 50\% due to lack of relevant counter information. The impact of DVFS on power modelling is not considered. Also targeting desktop GPUs, the authors of ~\cite{mian12} introduce a support vector regression model instead of the least squares linear regression more commonly used. A total of five variables are used, such as \textit{vertex shader busy} and \textit{texture busy} to build the model. Instead of predicting the power of full kernels as done in ~\cite{nag10}, they predict the power of the different execution phases as similarly done in our work. The authors show a slight advantage of SVR in accuracy although some phases of execution of the GPU power cannot be modelled correctly. The performance and power characterization done in  ~\cite{abe14} considers different desktop NVIDIA GPU families (i.e. Tesla, Fermi and Kepler). The external power measurements apply to the entire system which includes the GPU and CPU and not the individual components. The proposed power model uses performance counters and linear regression and introduces a frequency scaling parameter in the power equation to account for the different performance levels possible in the GPU. It does not consider the operating voltage and, with multiple voltage levels possible for a single frequency, this  could explain the errors in the prediction accuracy which are measured at around 20 to 30\%.  The work of ~\cite{Mei16} also focuses on desktop GPUs with  a review that shows that the number of explanatory variables used varies between 8 and 23. It considers the use of neural networks to perform the prediction, indicating how neural networks can address the nonlinear dependencies of the input variables at the expense of significantly  higher complexity. However, this could make the models harder to deploy as part of an energy-aware operating system. In this paper, we focus on using a low number of explanatory variables to make the models easy to deploy at run-time and investigate the accuracy of multiple linear regression for power modelling considering voltage, frequency and temperature in embedded GPUs.

\section{Methodology}

The methodology is based on our previous work targeting ARM big.LITTLE SoCs and introduced in  ~\cite{Nikov2020}. 
The CUDA benchmarks used for the model creation and validation are shown in Table ~\ref{tab:table2}. Training and testing benchmarks are independent and have been obtained from the Rodinia and CUDA SDK benchmark sets.

\begin{table}[h!]
  \begin{center}
    \caption{Deployed benchmarks}
    \vspace{5mm}
\begin{tabular}{ll}
    \label{tab:table2}
CUDA Rodinia Train Set     &            \\ \hline
stream\_cluster srad\_v1   & srad\_v1   \\
particle\_filter srad\_v2  & srad\_v2   \\
mmumergpu pathfinder       & pathfinder \\
leukocyte myocite          & myocite    \\
lavaMD kmeans              & kmeans     \\
backprop bfs               & bfs        \\
b+tree cfd                 & cfd        \\
heartwall hotspot3d        & hotspot3d  \\
hotspot hybridsort         & hybridsort \\ \hline
CUDA SDK Test Set          &            \\ \hline
binomialOptions Montecarlo &            \\
blackscholes particles     &            \\
SobolQRNG Radixsort        &            \\
Transpose FDTD3d           &            \\
Texture3D nbody            &            \\ \hline
\end{tabular}
  \end{center}
\end{table}

\begin{figure}[htbp]
\includegraphics[width=9cm,center]{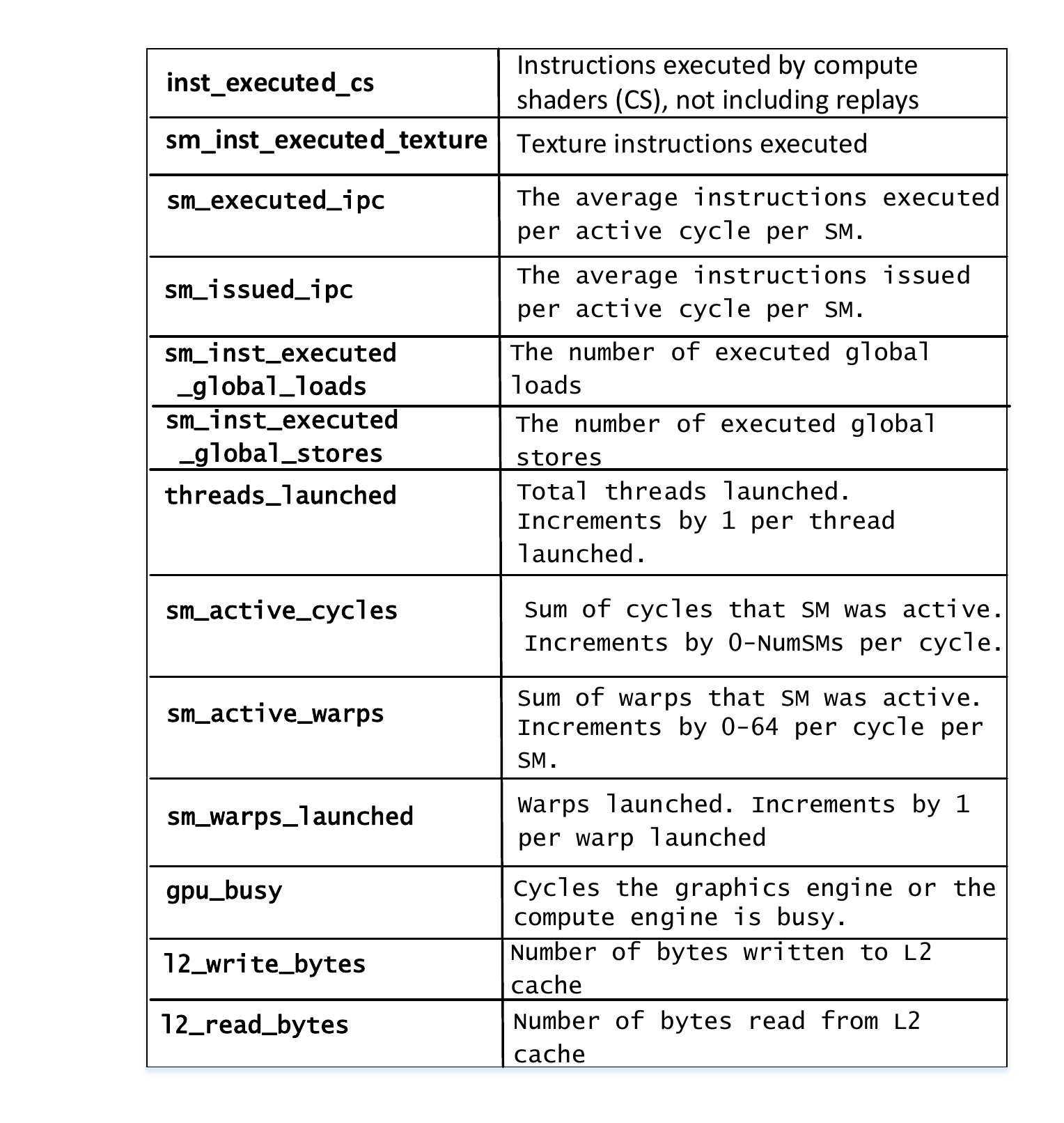}
\caption{Analyzed GPU performance counters}
\label{fig:pc}
\end{figure}

 We have modified the collection and processing stages to account for the differences in counter availability, power and current sensors and DVFS implementations. In the CPU-based power model done in ~\cite{Nikov2020} the DVFS table contains fixed pairs of voltage and frequency and the preferred way to build the power model is to use a per-frequency model in which a distinct set of coefficient values are calculated for each pair. To use this approach directly on the TX1 is problematic due to temperature dependencies and the practical difficulties of adjusting the temperature of the device to each possible value during a data collection run that typically executes over several days. To manage this complexity in this work we distinguish between local events such as the number of instructions executed or the number of memory accesses that will affect power in certain regions of the device, and global states such as the operating frequency, voltage and temperature that will affect power globally. This approach enables us to propose an unified model with a single set of coefficients that could be used for multiple combinations of voltage, frequency and temperature.  The development of this unified model and its comparison with the per-frequency models is conducted in section ~\ref{sec:um} and Section 7.  The performance counters considered in this work are shown in Figure~\ref{fig:pc}. The number of physical registers available in the GPU device to collect activity information in parallel is limited and for this reason limiting the number of model counters is preferred. The methodology presented in ~\cite{Nikov2020} implements different types of automatic searches and analysis of the effects of different counters on the power model accuracy.  

In the methodology flow, the octave\_makemodel script receives with -r a measurement.txt text file containing the power and activity counter samples with around 12,000 samples in our case. Then with -b a benchmark.txt file that identifies which benchmarks should be used for training and which for testing. Then with -f all the frequency values that are going to be considered (each frequency value also corresponds to a different voltage as determined by the DVFS table), -p identifies the column number in measurement.txt that contains power information, -m set to 1 is the search mode heuristic defined as bottom-up and -l lists the performance counters selected for analysis as columns in measurement.txt, -n set to 4 instructs the framework to search for the best possible four performance counters that result in a more accurate model as indicated with -c 1.  This means that the script will search up to a maximum of 4 performance counters in the list provided, across all the frequencies and voltages automatically. The result is a set of coefficients for each frequency/voltage pair. To minimize temperature interference, the experiments are conducted setting the available TX1 fan to maximum speed initially. If, for example, the user is interested in obtaining models across all possible frequencies for a particular set of events that have been pre-selected, the user can use the switch -e to specify the four columns in power\_measurement.txt with events that need to be analyzed. 


\section{per-frequency model development}
\label{sec:pf}

Equation ~\ref{eq:power_per_freq} shows the general form of the power model proposed in this work. Comparing this equation with the previous work done in ~\cite{Nikov2020} we normalize the total event count with the total number of cycles available in the time slot to obtain an activity density measurement that should remain constant as frequency changes. For example, if the frequency doubles then the number of events (e.g. instructions executed) in the same time period should also double, but since the number of clock cycles also doubles the ratio should remain constant. 



\begin{figure}[!htbp]
\includegraphics[width=9cm,center]{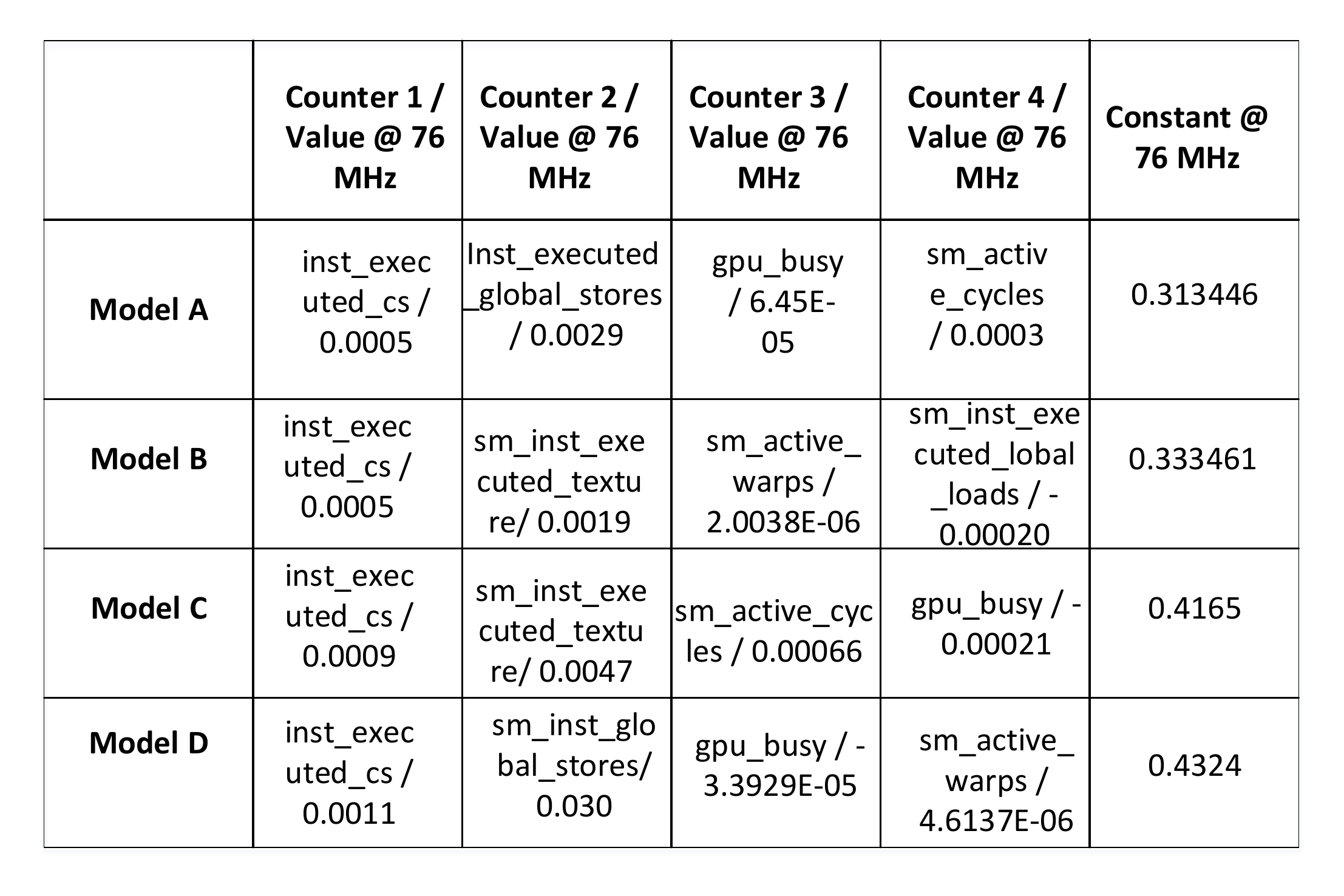}
\caption{Model parameters}
\label{fig:models}
\end{figure}

We limit all experiments to a maximum set of four counters to account for the limited number of registers available in commercial GPUs.
Figure ~\ref{fig:models} shows the four examples of performance counters that the methodology ends up selecting as the more accurate identified as model A, B, C and D. The coefficients shown are for a single example frequency of 76MHz with a corresponding voltage of 0.82v and a similar set of coefficients exists for the other 12 possible frequency and voltage pairs at a constant temperature.  

\begin{dmath}
P_{GPUfreq_1}= {\alpha}_0 + {\alpha}_1 \times events_1/cycles + \ldots + {\alpha}_n \times events_n/cycles
\label{eq:power_per_freq}
\end{dmath}

\begin{figure}[!htbp]
\includegraphics[width=10cm,center]{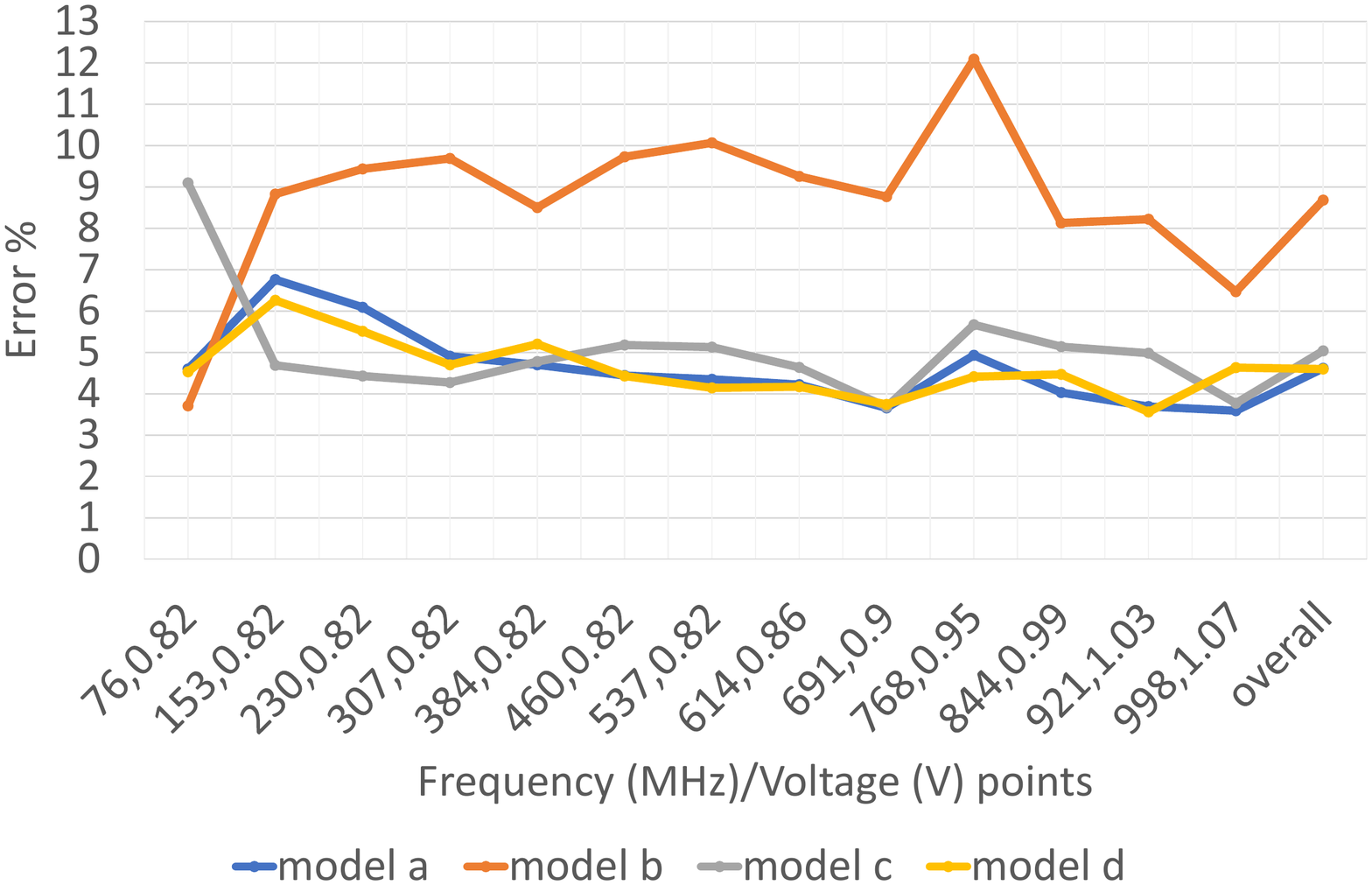}
\caption{Model comparison}
\label{fig:comp}
\end{figure}

Figure ~\ref{fig:comp} shows the comparison of the accuracy of these four models across all the frequency and voltage pairs. The performance of models A, C and D is similar with an overall error below 5\%. Model D offers a slightly  better overall accuracy, as shown in the overall value and will be taken forward to derive a unified power model in the next section. We can also appreciate that at different frequencies, the accuracy varies and this is largely defined by the model parameters.

\section{Unified model development}
\label{sec:um}

The previous per-frequency models contain a total of 13 times 5 parameters with four event coefficients and a constant parameter for each of the 13 voltage/frequency points. They are obtained at fixed voltage and frequency pairs and do not take into account the multiple voltage levels available for each frequency in the TX1 device part of the DVFS table. In this section, we propose a new type of model that unifies the previous models with a single set of coefficients and includes independent variables for frequency and voltage. Equation ~\ref{eq:power_um} shows the general form of this unified power model with two terms being added corresponding to dynamic and static power. The approach consists of scaling the power predicted by a power model at a single frequency to fit the rest of the frequencies and voltages. 

\begin{dmath}
P_{GPUfreq_x}= (P_{GPUfreq_1} - P_{GPUsta_x})\times \frac{freq_1}{freq_x} \times (\frac{volt_1}{volt_x})^2+P_{GPUsta_x}\times(\frac{volt_1}{volt_x})^2
\label{eq:power_um}
\end{dmath}

\begin{figure}[!htbp]
\includegraphics[width=10cm,center]{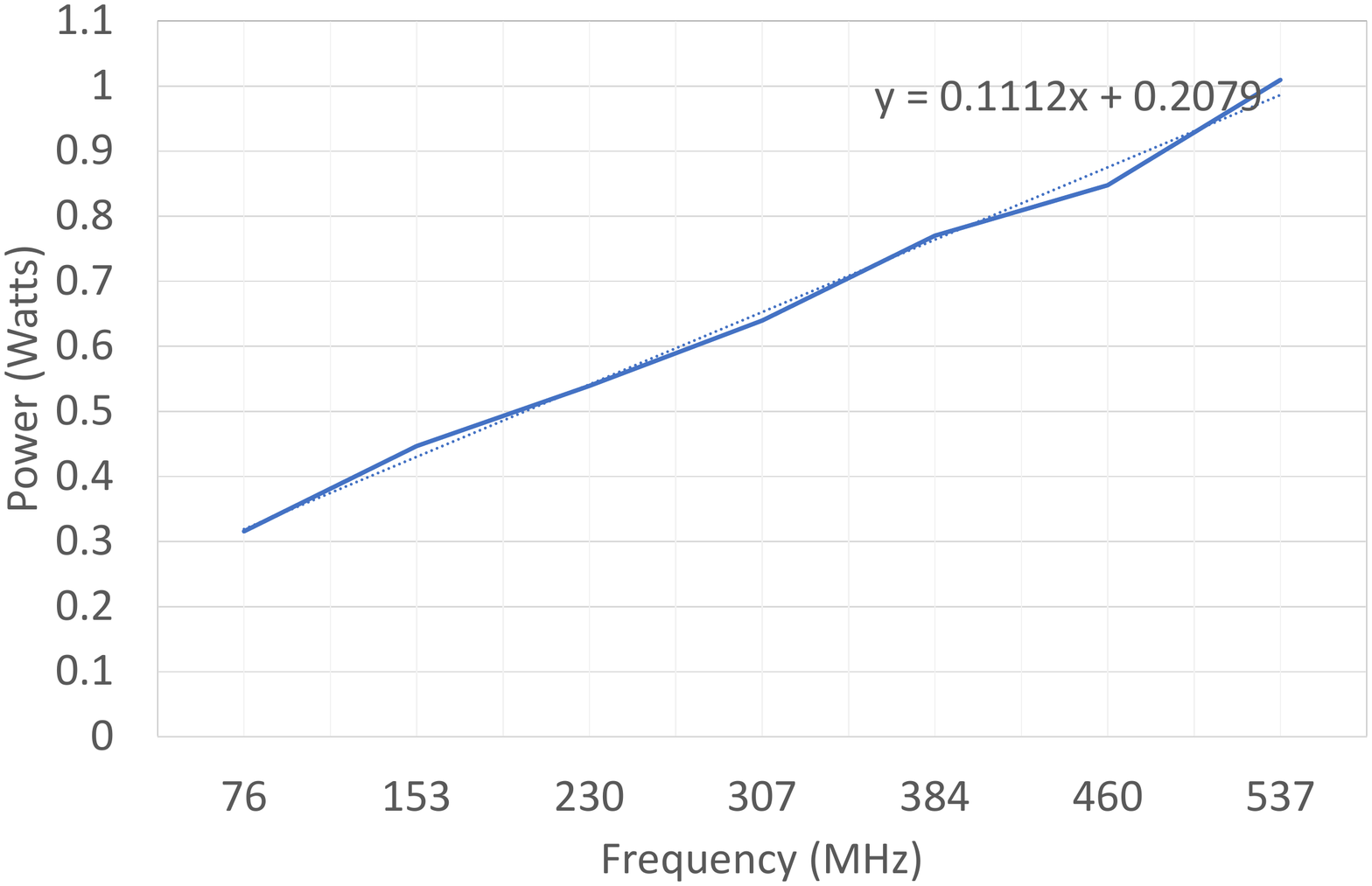}
\caption{Idle power estimation}
\label{fig:static}
\end{figure}

Scaling is possible because the model uses normalized activity rates that should remain constant at different frequencies since both events and cycles should reduce proportionally. The scaling is done based on how voltage and frequency affect the dynamic and static power of a chip. Dynamic power is proportional to the voltage square and frequency. In our experiments, we observe that static power accuracy improves using also voltage square scaling. Static power or leakage is the power of the device when the frequency is zero, so the frequency term should not be used to scale it. To isolate the static power in the second term of the equation to be able to scale it correctly, we need to measure it first. It is important to note that the per-frequency model contains a constant component that represents the device power with no activity and this power can be defined as idle power as shown in Equation ~\ref{eq:power_total}. This idle power is formed mainly by the static power and the clock power since the clocks remain active when there is no active load. A direct way to measure static power will be to clock gate the GPU device, however, the Linux for Tegra L4T JetPack 4.2.1 for the TX1 SoC used in this work does not implement this feature and only allows frequency configurations part of the DVFS table. To be able to extract the static power, we use an indirect method as follows. We sweep all the points available in the DVFS table with no benchmarks running to obtain the idle power. 
The first few frequency points in the DVFS table do not affect the supply voltage of 0.82V and this results in a linear relation of power and frequency as shown in Figure ~\ref{fig:static} for these points at a reference temperature of 23C at full fan speed. We use the point at which this line intersects the Y axis as the frequency of zero and the corresponding value rounded to 0.21W as the static power present in the device at that voltage level and temperature. With this information, we can create a unified model based on Equation ~\ref{eq:power_um} using as reference point any frequency that has the common voltage point of 0.82V and should have constant static power. 

\setlength\belowcaptionskip{-3ex}

We consider two possible reference points at the minimum frequency of 76 MHz and the middle frequency of 380 MHz that both share the voltage of 0.82V. We call these models UAL and UAM for "unified anchor low" and "unified anchor middle", respectively. Another alternative is to use a reference point at a high frequency if we can estimate the static power at that level.  Since we know that the dynamic power follows equation ~\ref{eq:power_dyn}, we can solve equation ~\ref{eq:power_total} and with the available values for \textit{P\textsubscript{idle}}, \textit{P\textsubscript{static}}, \textit{V} and \textit{f} we can obtain the \textalpha  \texttimes \textit{C} that we treat as a constant. Our hypothesis is that the activity rate \textalpha  \texttimes \textit{C} should remain constant within a small sample interval across different frequencies because we are measuring events divided by cycles in the sample interval. We can now extract the static power for the high frequency point of 998MHz and 1.07V by obtaining \textit{P\textsubscript{dynamic_clock}} and subtracting it from \textit{P\textsubscript{idle}}. We call this model UAH for "unified anchor high". 

\begin{figure}[!htbp]
\includegraphics[width=10cm,center]{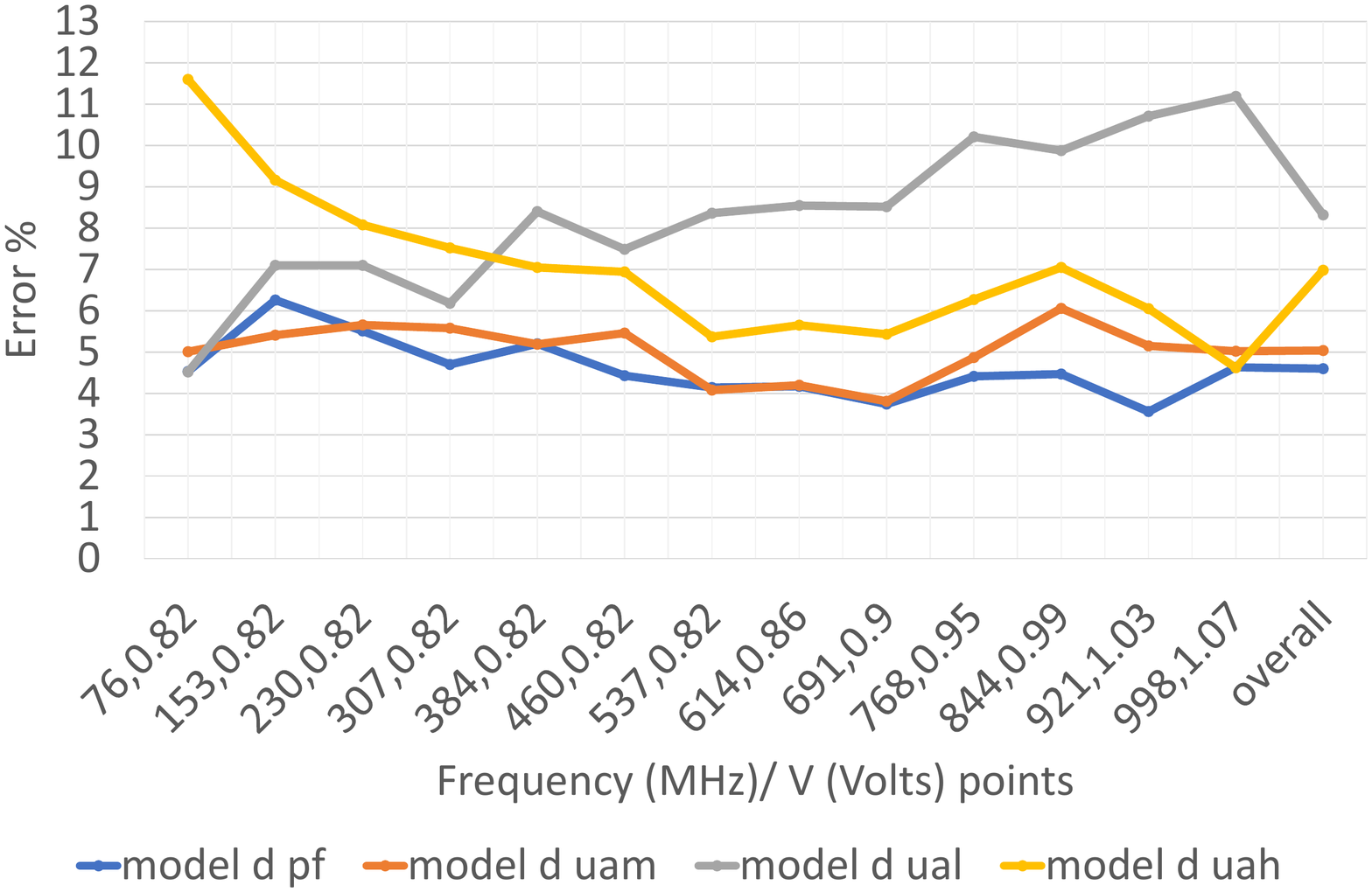}
\caption{Unified power model comparison}
\label{fig:anchor}
\end{figure}

Figure ~\ref{fig:anchor} shows the accuracy of the  UAL, UAM and UAH unified models. This figure shows how the accuracy compares between these unified models derived from the PF (per-frequency) model D. The anchor low and anchor high models obtain the best accuracy at their respective reference points but suffer a significant degradation as the frequency/voltage moves further away from the reference point. On the other hand, the anchor middle offers a largely identical accuracy to model D with an overall percentile error rate of around 5\%. This result shows that the unified model can be competitive in terms of accuracy compared with the per-frequency models developed in the previous section.  This unified power model with 380 MHz and 0.82V as the reference frequency and voltage is shown in Equation ~\ref{eq:power_umx} where \textit{P\textsubscript{GPUfreq_ref}} is obtained by Equation ~\ref{eq:power_ref}. Equation ~\ref{eq:power_ref} contains a negative coefficient which, in principle, is not an intuitive result but it can be explained by the fact that the different explanatory variables have correlations among them (i.e. GPU busy increases as the number of instructions executed increases) and the multiple linear regression process finds this negative value as a value that improves the model fit to the training data.

\begin{figure}
\includegraphics[width=10cm,center]{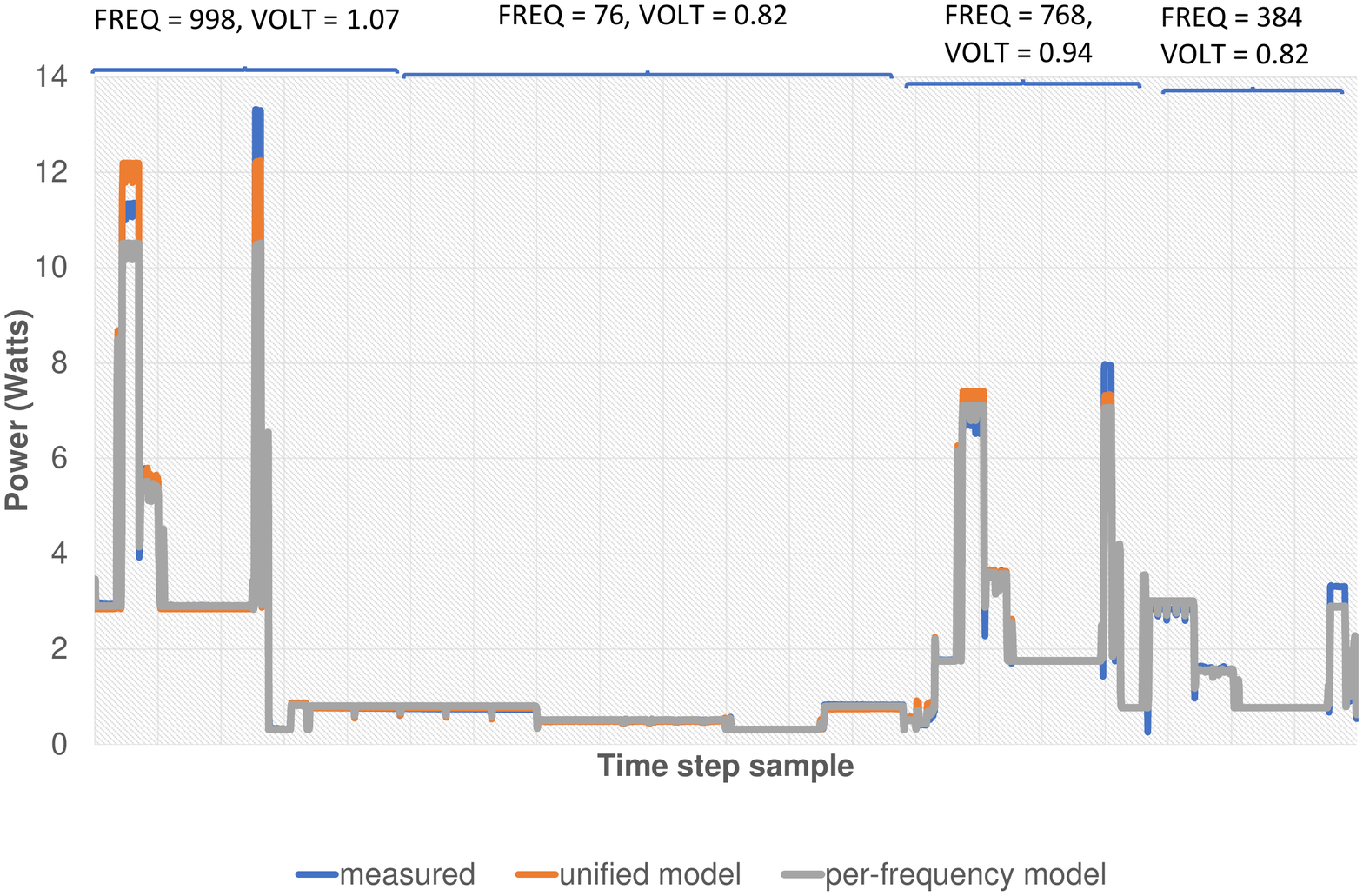}
\caption{Run-time prediction}
\label{fig:full}
\end{figure}

Finally, Figure ~\ref{fig:full} compares the power predictions performed by the per-frequency model D and the unified derived model with the measured values at run-time for a full sweep of the test benchmarks at different voltages and frequencies. We observe that the power consumption ranges from below 1 Watt to over 13 Watts depending on the operation point and benchmark. The power predictions follow the different execution phases, although it is at the highest points of power consumption that the errors are more noticeable. Also the measured power tends to show low spikes between benchmarks that the model does not predict. This effect could be due to our sampling frequency that is limited to one sample every 0.5 seconds. Further research is needed to increase this sample rate taking into account that, since the thread that samples the power sensors is also executed by the CPU cores higher sampling rates could mean that the processors are not available to launch the CUDA benchmarks which could introduce artifacts.

\begin{dmath}
P_{dynamic\_clock}= \alpha \times C \times V^2 \times f
\label{eq:power_dyn}
\end{dmath}

\begin{dmath}
P_{idle}= P_{dynamic\_clock} + P_{static}
\label{eq:power_total}
\end{dmath}

\begin{dmath}
P_{GPUfreq\_x}= (P_{GPUfreq\_ref} - 0.21w)\times \frac{freq\_x}{380MHz} \times (\frac{volt\_x}{0.82V})^2+0.21W\times(\frac{volt\_x}{0.82V})^2
\label{eq:power_umx}
\end{dmath}

\begin{dmath}
P_{GPUfreq\_ref}=  0.7720W + 0.0025W \times \frac{inst\_executed\_cs}{cycles} +  0.0908w \times  \frac{executed\_global\_stores}{cycles}  - 0.000017W \times  \frac{gpu\_busy}{cycles} + 0.000019W \times  \frac{active\_warps}{cycles}
\label{eq:power_ref}
\end{dmath}

\section{Temperature effects}

The power equation presented in Equation \ref{eq:power_umx} will consider possible changes in voltage and frequency due to temperature changes as defined in the DVFS table but it does not consider the changes in power due to temperature itself. Temperature has a direct effect on the static power consumption of the device as shown in \cite{GOEL20127}. The static power depends on leakage current and supply voltage linearly while the leakage current itself depends on the supply voltage and temperature exponentially. In this analysis we approximate these exponential relations linearly for the range in which device temperatures occur \cite{GOEL20127}.
To understand the dependency of static power and temperature we run a number of experiments with no load on the GPU varying the fan rate and frequency at a constant voltage of 0.82V to generate different temperature profiles. For each run we obtain a linear relation between frequency and power that enables as to estimate the static power by setting frequency at zero. We use the same approach to obtain a linear relation between power and temperature that allows to estimate the temperature at frequency zero for each of the runs. 

\begin{figure}[!htbp]
\includegraphics[width=10cm,center]{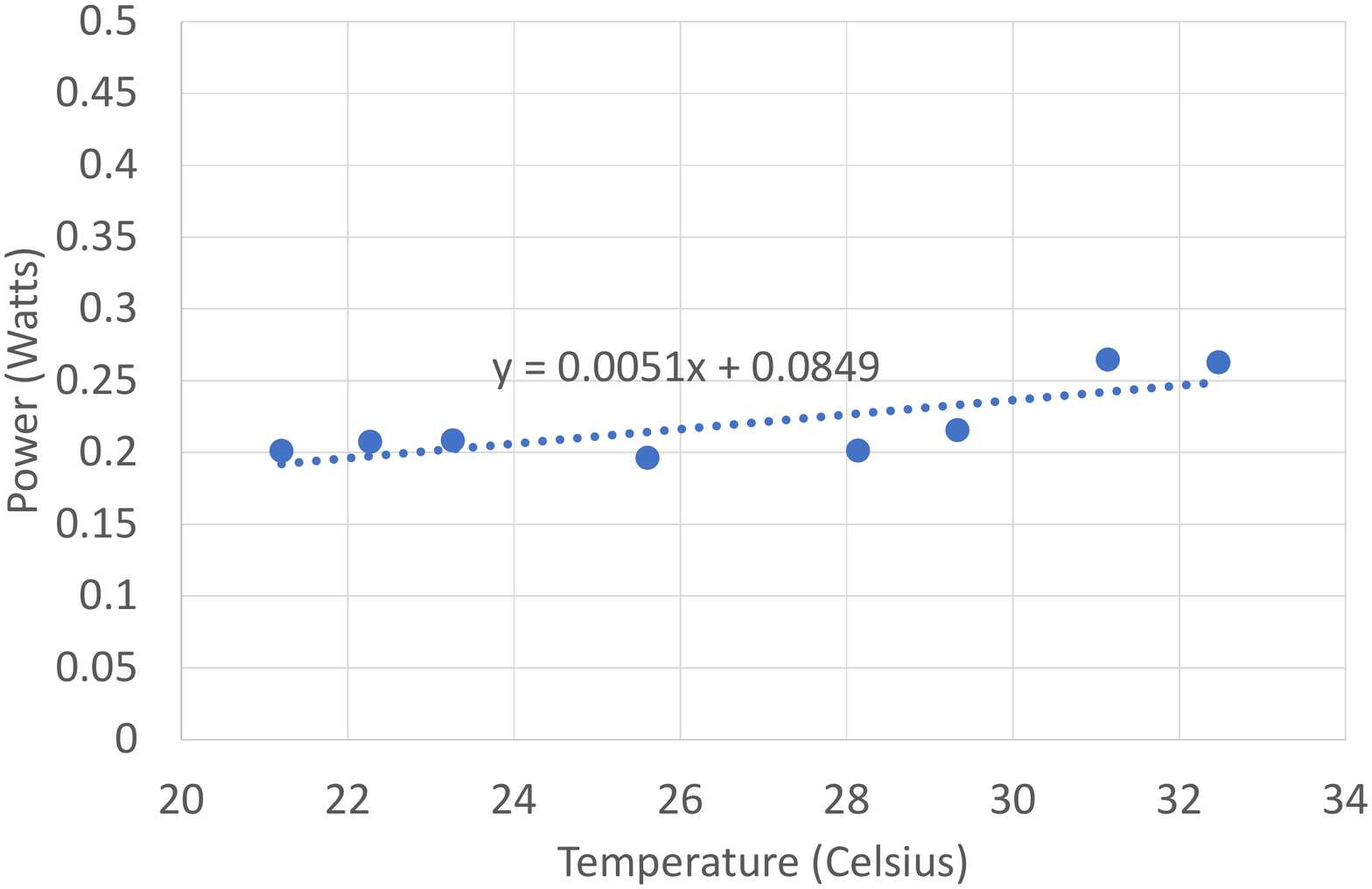}
\caption{Temperature and static power}
\label{fig:tempo}
\end{figure}

We can now plot the points of temperature and power as shown in Figure \ref{fig:tempo} and obtain a linear relation between temperature and static power at 0.82V. Using this information we can now replace the \textit{Pstatic} in Equation \ref{eq:power_umx} to obtain Equation \ref{eq:power_umxt}

\begin{figure}[!htbp]
\includegraphics[width=10cm,center]{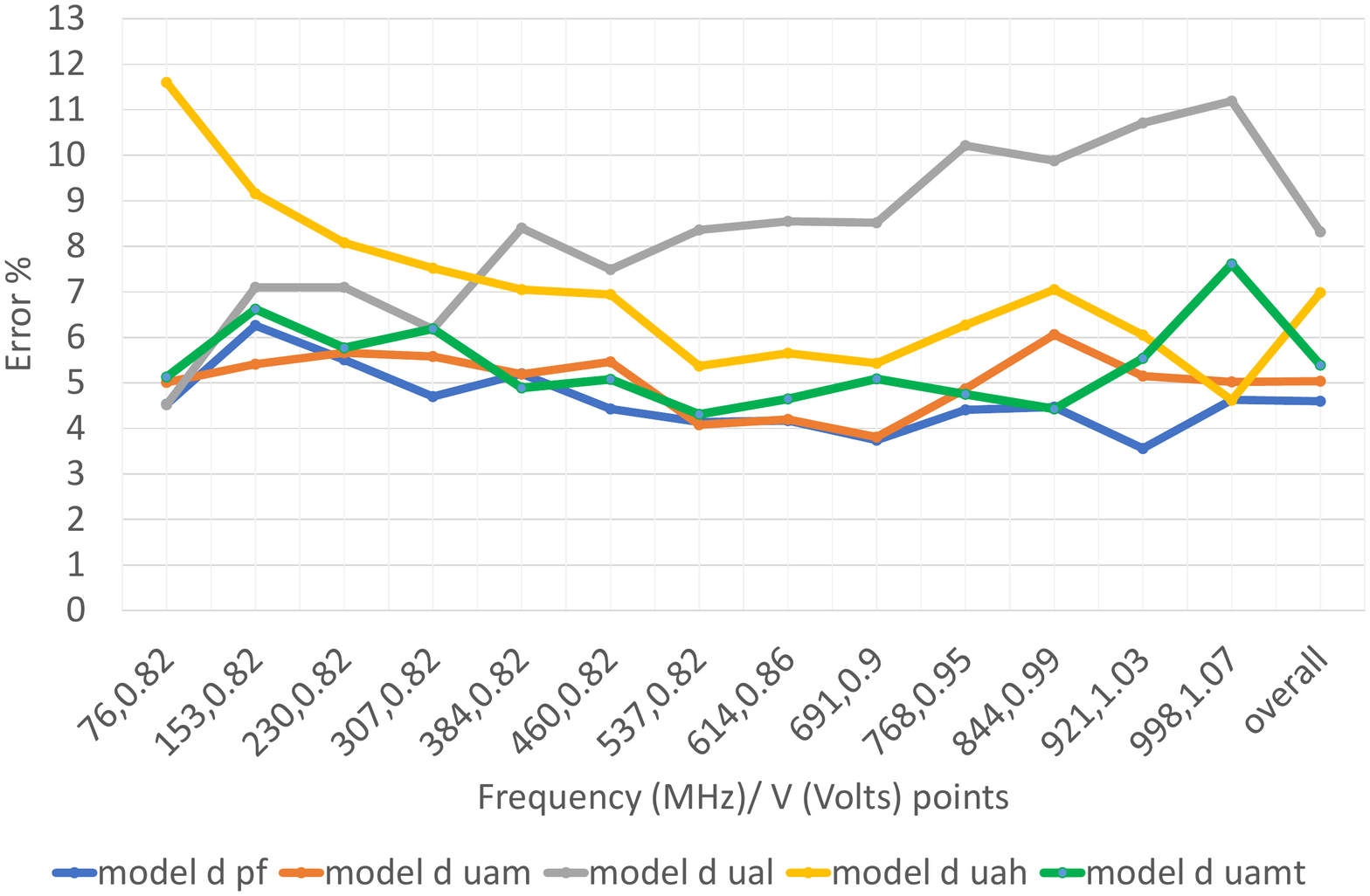}
\caption{Model comparison at low temperatures}
\label{fig:temp_255}
\end{figure}

\begin{figure}[!htbp]
\includegraphics[width=10cm,center]{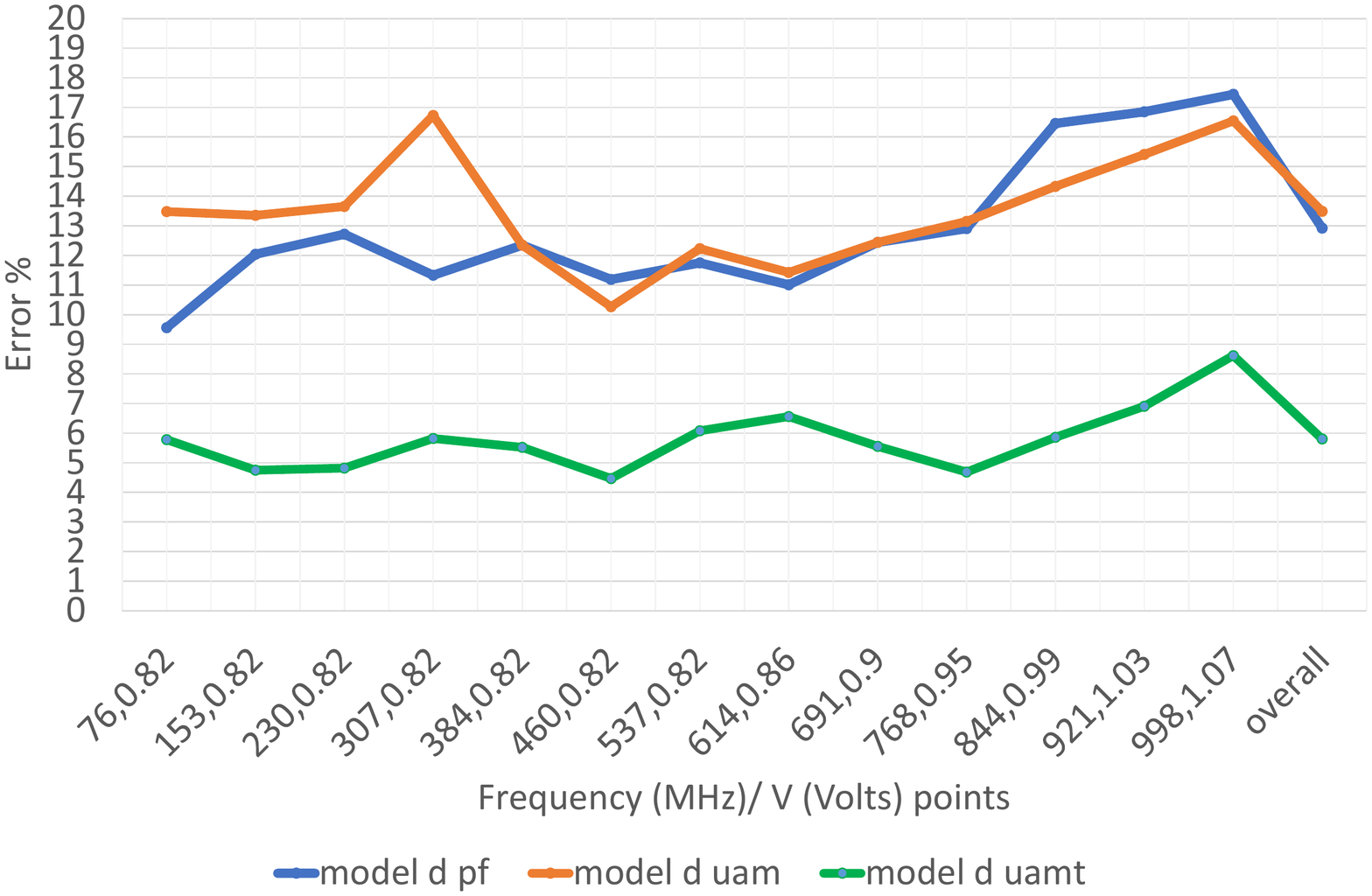}
\caption{Model comparison at high temperatures}
\label{fig:temp_0}
\end{figure}

Two temperature and power profiles generated by varying the fan activity are used to test the temperature-aware model. Overall the results show that under the same workload, voltage and frequency, temperature results in a power variation higher than 20\%. This accuracy result justifies the importance of capturing temperature in a power model as done in this work.  We evaluate the temperature-aware power equation in Figures \ref{fig:temp_255} and \ref{fig:temp_0} against the other models. Figure \ref{fig:temp_255} shows that under the same temperature conditions considered in the previous section, the model operates with a similar value of accuracy. Figure \ref{fig:temp_0} shows that when the device heats up the accuracy of the original models degrades significantly while the temperature-aware model largely maintains the same level of accuracy. 

\begin{dmath}
P_{GPUfreq\_x}= (P_{GPUfreq\_ref} -(T{ref}\times0.0051+0.0849)W)\times \frac{freq\_x}{380MHz} \times(\frac{volt\_x}{0.82V})^2+(T\times0.0051+0.0849)W\times(\frac{volt\_x}{0.82V})^2
\label{eq:power_umxt}
\end{dmath}

\section{conclusions and future work}

The proposed per-frequency and unified power models are kept simple by using only four performance counters collected in parallel. We also extend the unified model with temperature-aware capabilities to improve accuracy when the device can work in multiple temperature regimes such as in fanless configuration. We observe that the temperature impact on static power can increase power by over 20\% which is the rationale to make it a part of the proposed model. Overall, the research shows that the CPU power methodology can be applied successfully to GPU devices despite that the performance counters are very different in nature and that the prediction error can be maintained at around 5\% using a combination of local events represented by the performance counters and global states represented by voltage, frequency and temperature variables. The simplicity of these models means that they could be deployed as part of an energy-aware operating system and scheduling framework. The unified model could be particularly useful since it can capture multiple voltage levels for one frequency level with a single set of coefficients.  Our future work involves further validation of the methodology and its application with additional benchmarks, improving the data collection approach to increase the granularity of the samples to better capture the different phases of benchmark execution and experimenting with inter-prediction strategies across different GPU devices and technologies. The power modelling methodology used in this paper is available open-source at the following github repository ~\cite{buildmodel}.

\begin{acks}
This work was partially supported by the EPSRC ENEAC grant number EP/N002539/1, the H2020 TeamPlay project Grant agreement No.: 779882 and the Royal Society industrial fellowship MINET (Award: INF\char`\\R2\char`\\192044).
\end{acks}
%
%
%
\bibliographystyle{spmpsci}
%
\bibliography{ref}

\end{document}